\documentstyle[prl,aps,epsfig]{revtex}

\begin{document}


 \twocolumn[\hsize\textwidth\columnwidth\hsize  
 \csname @twocolumnfalse\endcsname              

\title{ Phase Transitions in Non-extensive Spin Systems}

\author{Robert Botet$^{\dagger }$, Marek P{\l }oszajczak$^{\ddagger }$ and 
Jorge A.
Gonz\'{a}lez$^{\star }$}
\address{$^{\dagger}$
Laboratoire de Physique des Solides - CNRS, B\^{a}timent 510, Universit\'{e}
Paris-Sud, Centre d'Orsay, F-91405 Orsay, France \\
$^{\ddagger}$
Grand Acc\'{e}l\'{e}rateur National d'Ions Lourds (GANIL), 
CEA/DSM -- CNRS/IN2P3, BP 55027, F-14076 Caen Cedex, France \\
and \\
$^{\star}$ Instituto Venezolano de Investigaciones Cient\'{i}ficas,
Centro de F\'{i}sica, Apartado 21827, Caracas 1020A, Venezuela}


\maketitle

\begin{abstract}
The spherical spin model with infinite-range ferromagnetic interactions is 
investigated analytically in the framework of non-extensive
thermostatics generalizing the Boltzmann-Gibbs statistical mechanics. 
We show that for repulsive correlations, a
new weak-ferromagnetic phase develops.
There is a tricritical point separating para, weak-ferro and ferro regimes.
The transition from paramagnetic to weak-ferromagnetic phase is 
an unusual first order phase transition in which a discontinuity of the 
averaged order parameter appears, even for finite number of spins. 
This result puts in a new way the question of the stability of critical
phenomena with respect to the long-ranged correlations.

\end{abstract}


 ]  

\narrowtext

Discussion of nonextensivity in thermodynamics go back to 70's \cite{mult8}.
Recently, non-extensive theories have become extremely important in
several areas of Physics \cite{mult1}. However, there are still very
important questions that should be addressed. The influence of the 
non-extensivity on phase transitions and their trace in finite systems 
is certainly a field which should be further explored \cite{nobre}. 
In the collisions of atomic nuclei or charged atomic clusters, a 
highly excited non-equilibrium transient system is formed which 
at a later stage of the reaction approaches an equilibrium and splits 
into many fragments \cite{mult3,mult7}. 
Observed signatures of criticality in these processes depend 
strongly not only on strong repulsive Coulomb interactions but also on
on the way the excited transient system is formed in collisions 
\cite{exp}.
The fragility of those signatures manifests the inherently 
non-extensive character of these 'critical phenomena' \cite{gppt00}. 
The general problem
of the relation between critical behavior in small
non-extensive statistical systems and the phase transitions in the
thermodynamical limit is investigated here using 
the Berlin-Kac Model (BKM) \cite{B&K}
in the framework of the Tsallis generalized statistical mechanics (TGSM) \cite{Tsallis}. 

The TGSM was inspired by earlier works on non-extensive thermodynamics, going back to the 60' \cite{mult8}, it
is based on an alternative definition for the
equilibrium entropy of a system whose $i$th microscopic state has probability
${\hat p}_i$ \cite{Tsallis} :
\begin{equation}
\label{eq1}
S_q=k\frac{1-\sum_{i}{\hat p}_{i}^{q}}{q-1}~~, 
\hspace{0.7cm} \sum_{i}{\hat p}_i=1~~, \hspace{0.7cm} k>0
\end{equation}
and $q$ (entropic index which is determined by the microscopic 
dynamics) defines a particular statistics. 
In the limit $q=1$, one
obtains the usual Boltzmann-Gibbs (BG) formulation of the statistical 
mechanics. The main difference between the BG formulation and the TGSM lies in
the nonadditivity of the entropy. 
For two independent subsystems 
$A$, $B$, such that the probability of $A+B$ is factorized into : 
$p_{A+B}=p_Ap_B$, the global entropy verifies :
$S_q(A+B)=S_q(A)+S_q(B)+(1-q)S_q(A)S_q(B) \ .$
The entropy $S_q$ has definite concavity for all
values of $q$. 
In particular, 
it is always concave for $q>0$, and
this is the case discussed in the present paper. 
It has been shown \cite{INVAR} that the TGSM
retains much of the
formal structure of the standard theory. Many important properties 
, such as the Legendre
transformation structure of thermodynamics and the H-theorem (macroscopic
time irreversibility) have been shown to be $q$-invariant. 
Considering the fact that the  essence of the
Second Law of Thermodynamics is concavity (see Ref. \cite{Lavenda}), 
the mentioned  properties of this entropy allow us to say 
that there are no problems with this law in TGSM.

The TGSM is relevant if the
effective microscopic interactions are long-ranged and/or the effective
microscopic memory is long-ranged and/or the geometry of the system is fractal 
\cite{Tsallis,Tsallis2,waldeer}. 
In the superadditive regime ($1-q>0$),
independent subsystems $A$ and $B$ will tend to join together 
increasing in this way the entropy of the whole system. On the contrary,
in the subadditive regime ($1-q<0$), the 
system increases its entropy by fragmenting into separate subsystems.
These ideas are in agreement with the results of Landsberg et al.
\cite{landsberg2,landsberg}
The subadditivity of entropy is expected 
when long-range repulsive interactions and correlations and/or long-range
memory effects 
are present in the system \cite{Tsallis2,landsberg}. This is in particular 
the relevant limit of the fragmentation of electrically charged 
(off-equilibrium) systems, 
such as formed in the collisions of atomic
nuclei or sodium clusters. 

The BKM was introduced 
as an approximation of Ising model. In the BKM, the individual spins are taken 
as
continuous 3-dimensional variables, but the rigid constraint on each local 
spin variable $S_i$
(where $S_i = \pm 1/2$) in Ising model, is 
relaxed and replaced by an overall spherical constraint :
\begin{eqnarray}  
\label{cons}
\sum_{i=1}^{i=N} S_i^2 = \frac{N}{4} 
\end{eqnarray}
for $N$ spins. In the present work, we
consider the infinite-range version of the BKM where all
pairs of spins interact with the same strength, according to the
Hamiltonian :
\begin{eqnarray}
\label{ham}
H = - \frac{J}{N} \sum_{(i,j)}S_i S_j -\frac{J}{8} ~ \ .
\end{eqnarray}
The sum in (\ref{ham}) runs over all different pairs of spin indices $(i,j)$. 
The constant term $-J/8$ is added to fix the origin of the energies at
the vanishing total magnetization : $\sum_i S_i =0$.
The normalization of the ferromagnetic ($J>0$) exchange energy as $J/N$, 
ensures finiteness of the energy per spin in an infinite system. The value
of this
energy should depend on the volume of the system.
 
One defines the magnetization per site by :
$\eta = N^{-1}\sum_i S_i ~ \ .$
Because of the inequality : $(\sum S_i)^2 < N\sum S_i^2$, the value of 
$\eta$ is contained in between $-1/2$ and $+1/2$. 
The phase of the system will be
characterized by the average value of $\eta^2$ : $m^2 = < \eta^2 > ~,$
where the brackets $<\cdots>$ denote the ensemble average at a given
temperature. 
The square root of $m^2$ plays the role of the order parameter.

The Hamiltonian of BKM can be expressed as a function of $\eta$ :
\begin{eqnarray}
\label{energy}
H = -\frac{JN}{2} \eta^2 ~ \ ,
\end{eqnarray}
and this allows us to obtain the degeneracy factor of the
macroscopic states defined by the fixed number of spins $N$ and the fixed
energy $E$. The number of such states in the 
$N$-dimensional spin-space is the volume 
of the intersection of the hypersphere (\ref{cons})
and of the hyperplane : $\sum_i S_i = ({-2NE/J})^{1/2}$. 
This intersection is a sphere of radius 
$(N/4+2E/J)^{1/2}$ in the $(N-1)-$ dimensional space and its 
volume is : $g_N(E)=a_N(1/4+2E/JN)^{N/2-1}$, where $a_N$ is
a numerical factor which depends only on the number of spins (system mass). 
Knowledge of this statistical weight,
in addition to the energy given by (\ref{energy})
allows us to compute all thermodynamic quantities of the equilibrated
system.

In BG statistical mechanics, the partition function is written 
in terms of the $\eta$-distribution as :
\begin{eqnarray}
Z = A_0 \int_0^{1/2} \eta ~~ \mbox{exp} (-Nf(\eta)) d\eta
\end{eqnarray}
The constant $A_0$ is : $2^{5-N}JNa_N \mbox{exp} (-\beta J/8)$ and : 
$$f(\eta) = -\beta J \eta^2 - (1-2/N) \mbox{ln} (1-4 \eta^2)/2$$ 
is the free energy per spin. 
In the limit $N \rightarrow \infty$, this function does not depend 
on $N$ which ensures the existence of the thermodynamic limit. 
Since this free energy is analytical in the variable $\eta$ : 
$f(\eta) = [4(1-2/N)-\beta J]\eta^2/2 + 8(1-2/N) \eta^4 + \cdots ~ \ ,$ 
with a positive $\eta^4$-term and a change in sign of the $\eta^2$-term at the
pseudo-critical point
$(\beta J)_{c,N} =4(1-2/N)$, the infinite system undergoes a second order phase
transition at the critical temperature $(\beta J)_c=4$. The critical
exponents are given by the regular Landau-Ginzburg mean-field theory for the
magnetic systems.

In the TGSM, the equilibrium energy distribution is given by \cite{Tsallis} :
\begin{eqnarray}  \label{pNq}
p_{N,q}(E) = \frac{(1-\beta (1-q) E)^{1/(1-q)}}{Z_q} ~ \ ,
\end{eqnarray}
if $1-\beta (1-q) E > 0$, and 0 otherwise. 
The partition function
$Z_q$ is defined by the normalization of $p_{N,q}$ :
\begin{eqnarray}
\int g_N(E) p_{N,q}(E) dE = 1  ~ \ .
\end{eqnarray}
All ensemble averages are then defined with respect to the properly 
normalized probability : 
$p_{N,q}^q(E) g_N(E)$~ \cite{Norm}. For example, the
averaged order parameter is given by :
\begin{eqnarray}
m_q^2 \equiv <\eta^2>_q = \frac{\int \eta^2(E) p_{N,q}^q(E) g_N(E) dE} {\int
p_{N,q}^q(E) g_N(E) dE} ~ \ .
\end{eqnarray}
Replacing the integration variable $E$ by $\eta$, the value of 
$m_q$ is :
\begin{eqnarray}  
\label{mq2}
m_q^2 = I_3/I_1 ~ \ ,
\end{eqnarray}
with the integrals $I_s$ defined by :
\begin{eqnarray}  
\label{Is}
I_s &=& \int_0^x \eta^s \left[ 1-\frac{\beta
J \kappa}{2}\eta^2 \right]^{-N/\kappa -1} (1-4\eta^2)^{N/2-1} d \eta  ~ \ ,
\end{eqnarray}
and $x \equiv {\mbox{min}(1/2,\sqrt{2/\beta J \kappa})}$ because of (\ref{pNq})
and $|\eta |<1/2$. The fundamental parameter of the non-extensive 
generalization of the BKM is : 
\begin{eqnarray}
\label{ooe}
\kappa = N(q-1) 
\end{eqnarray}
instead of $q-1$. This parameter, which will be called hereafter the 
Out-of-Extensivity (OE) parameter, allows us to define the thermodynamic 
limit of the system in a natural way. 
Let us compare systems of different masses $N$ such that the OE
parameter (\ref{ooe}) remains constant. In the relations (\ref{mq2}) and
(\ref{Is}), one can recognize the equivalent of a free energy per site :
\begin{eqnarray}
f_q(\eta) = \mbox{ln}(1-\beta J \kappa \eta^2/2)/\kappa - 
\mbox{ln}(1-4\eta^2)/2 ~ \ ,
\end{eqnarray}
which is independent of $N$ when $\kappa$ is a constant. 
Existence of this free energy per spin implies in turn
existence of the thermodynamic limit in the system.
For a fixed value of the OE parameter, the interesting values of $q$ approach 1
as the number of spins increases.
 
                  
When the number of spins $N$ is finite, 
the integrals $I_s$ can be expressed by the hypergeometric functions.
At high temperatures : $\beta J < 8/\kappa$, the upper bound for $\eta$ 
in (\ref{Is}) is $1/2$. The integrand in this case is
always finite, and $m_q$ is a positive decreasing
function of the temperature. Moreover, if : $\kappa >2/(1-2/N)$, 
then for all these temperatures one has : $\beta J <4 (1-2/N)$ and 
$m_q \propto 1/\sqrt N$, similarly to the standard paramagnetic phase. 
In particular, when $\beta J \rightarrow
8/\kappa$ by lower values then the limiting value is
$m_q=N^{-1/2}[\kappa/(2(\kappa-2))]^{1/2}$.
On the other hand if $\kappa <2/(1-2/N)$,  
then this $1/\sqrt N$-behavior of $m_q$ holds only
when $\beta J < 4(1-2/N)$. For higher values of $\beta J$, 
the leading behaviour of the order parameter
takes finite values independent of $N$, and tends to 1/2 when 
$\beta J \rightarrow
8/\kappa$ by the lower values.
\begin{figure}[t]
\epsfig{figure=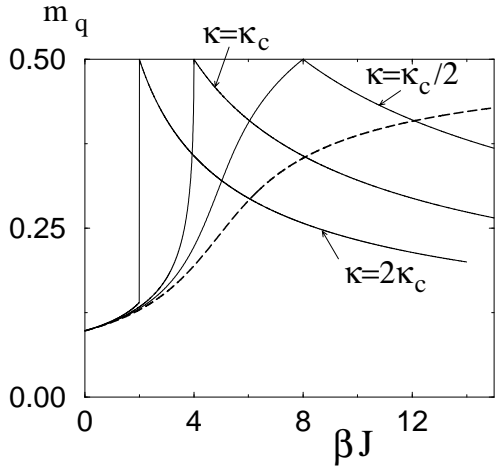,height=6cm}
\caption{Plots of the averaged magnetization of the infinite-range BKM 
in the TGSM for $N=50$ interacting spins 
and $\kappa=2\kappa_c$, $\kappa_c$, $\kappa_c/2$. 
$\kappa_c$ is the value of the OE parameter 
for which the pseudo-critical point of the
standard second order phase transition (here $(\beta J)_{c,N} = 3.84$) 
and the new critical point for the non-extensive BKM
($(\beta J)_{c,q} =8/\kappa$) occur at the same temperature. The BG limit
$(\kappa=0)$ is shown by the dashed line.}
\end{figure}

At low temperatures : $\beta J > 8/\kappa$, the quantities 
$[1-(\beta J \kappa/2)\eta^2]^{-N/\kappa -1}$ diverge for 
finite values of $\eta$ smaller than $1/2$, and the value of $m_q$ is
dominated by the behavior of the integrand near this divergency. In this case
: $m_q = \sqrt{2/(\beta J \kappa)} ~ \ ,$
since the order-parameter probability distribution collapses into the two Dirac
distributions.  Moreover, if $\beta J$ tends to
$8/\kappa$ by larger values, the limit of $m_q$-values  is 1/2. 
This implies that if $\kappa >2/(1-2/N)$, then the averaged order
parameter $m_q$ has a discontinuous jump at the
critical temperature $(\beta J)_{c,q} = 8/\kappa$ between the small value of 
order $1/\sqrt N$ and 1/2. Otherwise, if $\kappa < 2/(1-2/N)$, then $m_q$
is continuous but 
its first temperature derivative becomes discontinuous 
at this critical temperature 
$(\beta J)_{c,q} = 8/\kappa$, leading 
to the second order critical phenomenon. Both behaviors are exemplified in 
Fig. 1. The tricritical point ($(\beta J)_{TC} = 4$, $\kappa_{TC}=2$) separates the 
phase boundary line in transitions of different nature 
(discontinuous/continuous).
One of the continuous phase transitions is the simple continuation of the
well known paramagnetic $\Leftrightarrow$ ferromagnetic transition 
for the BG statistics, while the other one is
a new second order phase transition (ferromagnetic $\Leftrightarrow$
weak-ferromagnetic) due to the presence of the long-range 
repulsive correlations/memory effects.   
\begin{figure}[h]
\epsfig{figure=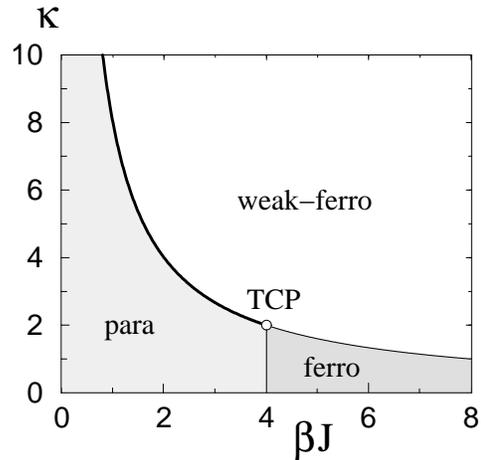,height=6cm}
\caption{
Phase diagram of the infinite-ranged BKM ferromagnetic spin model in the
TGSM at the thermodynamic limit. The bold-face solid curve shows 
the line where the first order phase transition between
paramagnetic and weak-ferromagnetic phase happens. 
For $0<\kappa<2$, the
system undergoes two distinct second order phase transitions :
paramagnetic$\Leftrightarrow$ferromagnetic and 
ferromagnetic$\Leftrightarrow$weak-ferromagnetic. The tricritical point
is located in $(\beta J)_{TC} = 4$, $\kappa_{TC} = 2$.}
\end{figure}
Since the above limiting temperatures and OE parameters 
do not depend on the mass $N$ (the number of spins), the pattern of different 
behaviors of the order-parameter remains unchanged in the thermodynamic
limit ($N \rightarrow \infty$) and one may draw the phase diagram $\beta J
- \kappa$ ($\kappa > 0$) (see Fig. 2). 
The first order phase transition in the generalized BKM 
(the bold-face solid line in Fig. 2) 
remains unaltered in small systems, unlike the case of regular collective 
critical phenomena. The additional phase
appearing on this diagram is due to the long-range non-extensive 
correlations which tend to disorganize spin coherence.
This phase is called the weak-ferromagnetic state, 
because the order parameter in this state has a positive value,
but this value is lowered by the disruptive 
long-range correlations (see Fig. 1 in the case 
$\kappa = \kappa_c/2$). This is the reason why the magnetization in the
weak-ferromagnetic phase decreases when the
temperature is lowered.  

As we have seen above, the energy at constant volume in the BKM can be
written as a function of the squared order parameter. This allows us to
write the averaged energy as : $U_q = -J N m_q^2/2 \ .$ The specific
heat at constant mass and constant volume 
is the derivative of $U_q$ with respect to the temperature,
keeping the mass $N$ and the coupling $J$ fixed :
\begin{eqnarray}
\frac{C_q}{k_B} = \frac{(\beta J)^2}{2} N \frac{\partial m_q^2}{\partial
(\beta J)}
\end{eqnarray}
>From the above discussion, it is clear that there is a finite discontinuity 
of the specific heat when $\kappa < 2/(1-2/N)$.
But, there exists also a true divergence
of $C_q$ for the temperature $(\beta J)_{c,q} = 8/\kappa$ when $\kappa
\geq 2/(1-2/N)$, even if the system is finite. 
\begin{figure}[t]
\epsfig{figure=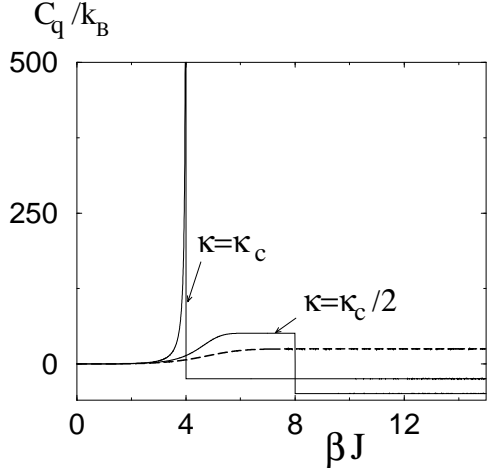,height=6cm}
\caption{
Plots of the specific heat in the infinite-range BKM ($N=50$) for : 
$\kappa = \kappa_c$ 
(the divergence at $(\beta J)_{c,q} = 8/\kappa$), and $\kappa = \kappa_c/2$ 
(the jump at $(\beta J)_{c,q} = 8/\kappa$). The BG limit ($\kappa = 0$) is
shown by the dashed line.}
\end{figure}

There is some evidence \cite{jund} supporting the fact that long-range
attractive interactions lead to $q<1$. We believe it is very important to
stress that long-range repulsive interactions can lead to non-extensivity
with $q>1$. 
The fact that the present system undergoes two different second-order phase
transitions in the region $q>1$ is very remarkable. On the other hand, in
this system the fundamental parameter is $N(q-1)$ instead of $q$. This is
a new result in non-extensive physics.

In conclusion, we have shown that the spherical
spin model with long-range correlations/memory effects simulated in the 
framework of non-extensive thermostatistics, develops 
a new weak-ferromagnetic phase  in a subadditive entropy 
regime. An unusual first order phase transition, which exhibits
discontinuity of the order parameter even in finite systems, separates
this phase from the standard paramagnetic phase. On the phase boundary
line in the plane $\beta J - \kappa$, we have found the tricritical point
separating the nature (discontinuous/continuous) of the transition.
Above a critical value of the OE-parameter, the spin system freezes
into the weakly ordered (weak-ferromagnetic) phase, passing through 
the phase boundary in the discontinuous transition. For a fixed value of the
OE-parameter, the thermodynamic limit ($N\rightarrow \infty$, $q-1 \rightarrow
0_+$) of the spin system differs from its BG limit. The non-extensivity
shields the system from the continuous disorder$\Longleftrightarrow$order
phase transition and suggests that, in general,  the second order 
critical phenomena may be unstable
in the presence of long-range repulsive correlations. This result
puts in a different perspective the 
discussion of a 'critical behavior' in collisions of
atomic nuclei or atomic clusters, showing
that the observed signals correspond possibly to a different 
limiting behavior than previously supposed. 

\vspace{0.2cm}
\noindent
{\bf Acknowledgements}\\
We thank K.K. Gudima for stimulating discussions.

\end{document}